\documentclass[pre,preprint,superscriptaddress,showpacs]{revtex4-1}
\usepackage{graphicx}
\usepackage{amsmath}
\usepackage{amssymb}
\usepackage{ulem}
\usepackage[usenames]{color}

\newcommand\st{\bgroup\markoverwith
{\textcolor{blue}{\rule[0.5ex]{2pt}{0.4pt}}}\ULon}

\begin{document}
\title{Cluster-size heterogeneity in the two-dimensional Ising model}

\author{Woo Seong Jo}
\affiliation{BK21 Physics Research Division and Department of Physics, Sungkyunkwan University, Suwon 440-746, Korea}
\author{Su Do Yi}
\affiliation{BK21 Physics Research Division and Department of Physics, Sungkyunkwan University, Suwon 440-746, Korea}
\author{Seung Ki Baek}
\affiliation{School of Physics, Korea Institute for Advanced Study, Seoul 130-722, Korea}
\author{Beom Jun Kim}
\email[Corresponding author, E-mail: ]{beomjun@skku.edu}
\affiliation{BK21 Physics Research Division and Department of Physics, Sungkyunkwan University, Suwon 440-746, Korea}


\begin{abstract}
We numerically investigate the heterogeneity in cluster sizes in 
the two-dimensional Ising model and verify its scaling 
form recently proposed in the context of percolation problems 
[Phys. Rev. E {\bf 84}, 010101(R) (2011)]. 
The scaling exponents obtained via the finite-size scaling analysis
are shown to be consistent with theoretical values of the fractal 
dimension $d_f$ and the Fisher exponent $\tau$ for the cluster distribution.
We also point out that strong finite-size effects exist due to the
geometric nature of the cluster-size heterogeneity.
\end{abstract}


\maketitle


In studies of critical phenomena,
the finite-size scaling (FSS) has proved to be an extremely fruitful
approach, owing to the recent development in computation~\cite{newmanMC}.
It is through a crossover effect that
this method connects numerical results to theoretical
understanding so successfully. Although a typical simulation
of a lattice system can only deal with a finite length scale $L$, one may
regard $L$ as a relevant scaling variable in the following
way~\cite{cardy}: Under a scaling transformation by
zooming factor $b$, i.e., 
$L^{-1} \rightarrow bL^{-1} \rightarrow b^2L^{-1} \cdots \rightarrow b^n
L^{-1}$, the scale invariance
of critical phenomena suggests that the singular part of the free energy $f$
will transform as
\begin{equation}
f(u_t, L^{-1}) = b^{-d} f(b^{y_t} u_t, bL^{-1}) = b^{-nd} f(b^{ny_t} u_t,
b^n L^{-1}),
\label{eq:free}
\end{equation}
where $u_t$ is a thermal scaling variable and $y_t = 1/\nu$ is its eigenvalue.
The quantity $\nu$ is called the correlation-length exponent since the
correlation length $\xi$ behaves as $\xi \sim |t|^{-\nu}$ with $t \equiv
(T-T_c)/T_c$, where $T_c$ denotes the critical temperature.
The scaling variable $u_t$ is proportional to $t/t_0$
near a fixed point of this renormalization-group (RG) transformation, where
$t_0$
is a nonuniversal scale specific to the system. This linear approximation
remains valid until the system reaches a point where $b^{ny_t}|t/t_0| \sim
\mathcal{O}(1)$. Comparing this with Eq.~(\ref{eq:free}), we find that $f$
is only weakly dependent on the first argument in this region and thus can
be written as
\begin{equation}
f(u_t, L^{-1}) \sim \left| t/t_0 \right|^{d\nu} f \left( |t/t_0|^{-\nu}
L^{-1} \right).
\label{eq:free2}
\end{equation}
In other words, the crossover occurs when $|t| \sim L^{-1/\nu}$, and
therefore, $L$ usually enters an FSS form accompanied by the exponent $-1/\nu$.
Recalling that
$\xi \sim |t|^{-\nu}$, we see that the argument on the
right-hand side of Eq.~(\ref{eq:free2}) means
$\xi/L$, expressing the competition between $\xi$ and $L$.

The cluster-size heterogeneity 
$H$ suggested by Lee {\it et al.} has been
devised in the context of recent debates on explosive percolation and
defined as the number of distinct cluster sizes~\cite{H.K.Lee}.
For brevity, we will call it simply ``the heterogeneity'' henceforth, but
it is to be noted that only sizes (or volumes) of clusters matter 
in calculation of $H$, irrespective of shapes of clusters.
Noh {\it et al.} have shown that this quantity can be applied to
ordinary percolation as well~\cite{Noh}:
They have calculated $H = H(p)$ with finite $L$'s, where $p$ means occupation
probability, and found that $H(p)$ has a peak at $p=p^\ast$,
which approaches the true percolation threshold $p_c$ as $L$
increases.
It is a typical signature to detect the percolation transition. One may
well expect that the deviation of $p^\ast$ from the true $p_c$, related to
the thermal scaling variable in percolation, will
be described as $\sim L^{-1/\nu}$ for the same reason as explained above.
However, Noh {\it et al.}~\cite{Noh} have
revealed that the FSS form for $H$ is
obtained with another exponent $\nu_H$ than $\nu$.
This exponent $\nu_H$ can be argued in the following way: Let us consider
the probability distribution of cluster sizes $s$, which scales as
\[ P(s) \sim s^{-\tau} e^{-s/s_c},\]
with the characteristic cluster size $s_c \sim |t|^{-1/\sigma}$.
Here $t = p-p_c$ in percolation problem and the scaling exponent 
$\tau$ is called the Fisher exponent.
If the system is off-critical, its cluster sizes will be simply found
between $1$ and $s_c$ so that $H \sim s_c \sim |t|^{-1/\sigma}$. On the
other hand, if $|t| \ll 1$, $H$ is limited by the finite size $L$ and its
scaling relation becomes different: We first consider $s_n$ such that
$P(s_n) \sim s_n^{-\tau} \sim 1/N_c$, where $N_c$ is the total number of
clusters. In fact, since the typical cluster size is
$\sum s P(s) \sim \mathcal{O}(1)$, $N_c$ is
comparable to the total number of points $N$. For a $d$-dimensional system,
it is obvious that $L$ is related to $N$ by $N \sim L^d$.
Furthermore, since $P(s_n) \sim
1/N_c$, there are few clusters above $s_n$ and their contribution to $H$ will
be much smaller than those with $s<s_n$. It is likely to find at least one
cluster for each $s<s_n$ since $P(s<s_n) > 1/N_c$, and thus it is plausible
that $H \sim s_n \sim N^{1/\tau}$. To sum up, $H$ is expected to have the
following behavior
\[
H(t,L) \rightarrow \left\{
\begin{array}{ll}
L^{d/\tau} & \mbox{if~~}|t|\ll 1\\
|t|^{-1/\sigma} & \mbox{otherwise.}
\end{array}
\right.
\]
Note that it is not $\xi$ and $L$ that actually compete here.
The competition is rather between $s_c$ and $s_n$, reflecting the limitation
in observing large clusters due to the finite system size.
The appropriate FSS form for describing $H$ should be, therefore,
\begin{equation}
H(t,L) = L^{d/\tau} F_H \left(|t| L^{1/\nu_H} \right)
\label{eq:hfss}
\end{equation}
with $\nu_H \equiv \tau/(d\sigma)$. This FSS form is readily supported by
numerical results in percolation~\cite{Noh}.

The ferromagnetic Ising model is one of the simplest and the most
important models in statistical mechanics. When the dimensionality $d$ is 
higher than unity, it undergoes an order-disorder transition at a critical
temperature $T_c$. Near the critical temperature
$T_c$, the magnetic order
parameter $m$ scales as a power-law form
$m \sim \left | t \right |^\beta$ and
the corresponding FSS form is given as $ m = L^{-\beta/\nu}
F_m(|t|L^{1/\nu})$~\cite{goldenfeld,newmanMC}.
Instead of this conventional order parameter,
we will look into the system in terms of cluster statistics such as
heterogeneity in the present study, where
a cluster in the Ising model is defined as a set of nearest-neighboring spins
with the same direction.
We first recall that
Suzuki~\cite{suzuki} has conjectured that the fractal dimension of an Ising
cluster is $d_f = d - \beta/\nu$.
For instance, for $d=2$, this conjecture suggests $d_f = 15/8$ by
inserting $\beta = 1/8$ and $\nu = 1$ into the relation. However, this
suggested fractal dimension is not consistent with
numerical results~\cite{cambier}.
The reason is that there are actually
two contributions in forming a cluster: One is
due to the correlation due to the spin interaction, while the other is a purely
geometric contribution which survives even in the infinite-$T$
limit~\cite{conig}. For example, one has a chance to find a giant
cluster in the Ising model in the triangular lattice at $T\rightarrow
\infty$ since each site has a spin state of either up or down randomly with
probability $1/2$, which coincides with the site-percolation threshold
in the triangular lattice~\cite{nappi}. In order to separate the geometric
effect from the correlation effect, one may consider bond-occupation
probability $p_B$ inside a cluster~\cite{conig,peruggi,van}. These bonds are
only to define connectivity between neighboring spins in the same direction
and do not affect the spin interaction.
Therefore, we are back to the original cluster statistics with
$p_B=1$, which is of our primary concern.
When $p_B=0$, on the other hand, the system reduces to the
pure Ising model, with two relevant eigenvalues associated with $t$ and
$h$, respectively, where $h$ is an external magnetic field.
It means that we need think of the RG
parameter space as $(t, h, p_B)$.
Since $p_B$ is related to the random bond percolation, the $p_B$-axis will
have one unstable fixed point, separating two stable fixed points. A careful
RG analysis shows that the stable fixed point at higher $p_B$ is actually a
tricritical point~\cite{stella,van}. Suzuki's argument that $d_f = 15/8$
is indeed true at the unstable fixed point describing Fortuin-Kasteleyn
clusters~\cite{fk}, but one should note that the
critical behavior is not given by this point, because it is the
tricritical point that attracts the RG flow starting from $p_B=1$.
Conformal invariance then predicts
that the fractal dimension for geometrical clusters of the
two-dimensional (2D) Ising model is $d_f = 187/96$ from
tricritical exponents~\cite{stella}.
This prediction is well substantiated by numerical results in
Refs.~\cite{Fortunato,Janke,frank}.

Now we consider the cluster-size distribution in the square lattice
at the critical temperature
$T_c = 2/\ln(1+\sqrt{2}) \approx 2.27$ in units of $J/k_B$, where $k_B$ is
the Boltzmann constant.
This distribution follows a power law with exponent $\tau$ in the
thermodynamic limit, and
the exponent $\tau$ follows the relation~\cite{christensen},
\begin{equation}
\tau = 1 + d/d_f,
\label{eq:tau}
\end{equation}
which is derived from cluster statistics as in percolation. In other words,
one can check the predictions of $d_f$ by measuring $\tau$ from the
cluster-size distribution.

We perform Monte Carlo (MC) simulations of the Ising model using the Metropolis
and the Wolff algorithms~\cite{newmanMC} in 2D $L \times L$ square with 
under the periodic-boundary condition. For most simulations we use 
$L=20, 40, 60, 80, 100$ and $160$, but $L=320$ is also used when required.
We mostly use the Wolff algorithm for efficiency, and the Metropolis algorithm 
only for a consistency check.
We start from a temperature much higher than $T_c$, 
and then slowly decrease $T$, measuring equilibrium quantities at each temperature.
All results are obtained from averages over $ 5 \times 10^5 $ MC steps, after disregarding the
first $5 \times 10^5$ MC steps for equilibration. 
We first choose a snapshot of a spin configuration in equilibrium and
identify all the clusters in the system. Then we count how many
different sizes of the clusters are found in spin-up and down directions,
respectively. 
The cluster heterogeneity is then calculated as the sum of these two numbers,
one in the spin-up direction, the other in the spin-down direction.  Our
brute-force approach in counting all the cluster sizes works as a main
bottleneck in increasing the system size: For example, the CPU time spent for $L=160$
roughly amounts to $4\times 10^2$ hours. Although limited by such a practical 
difficulty, our results nevertheless nicely agree with the prediction 
in the conformal field theory (see Table~\ref{table:exp}).


\begin{figure}
\includegraphics[width=0.42\textwidth]{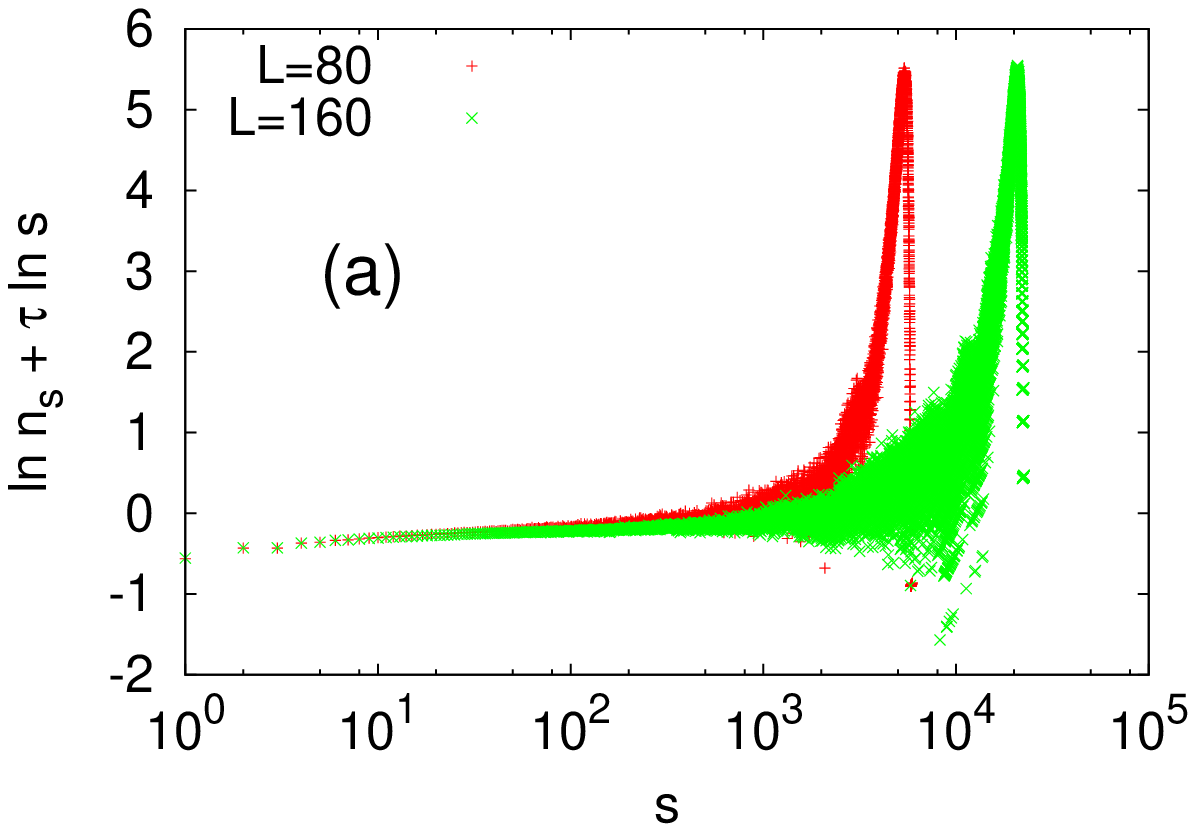}
\includegraphics[width=0.42\textwidth]{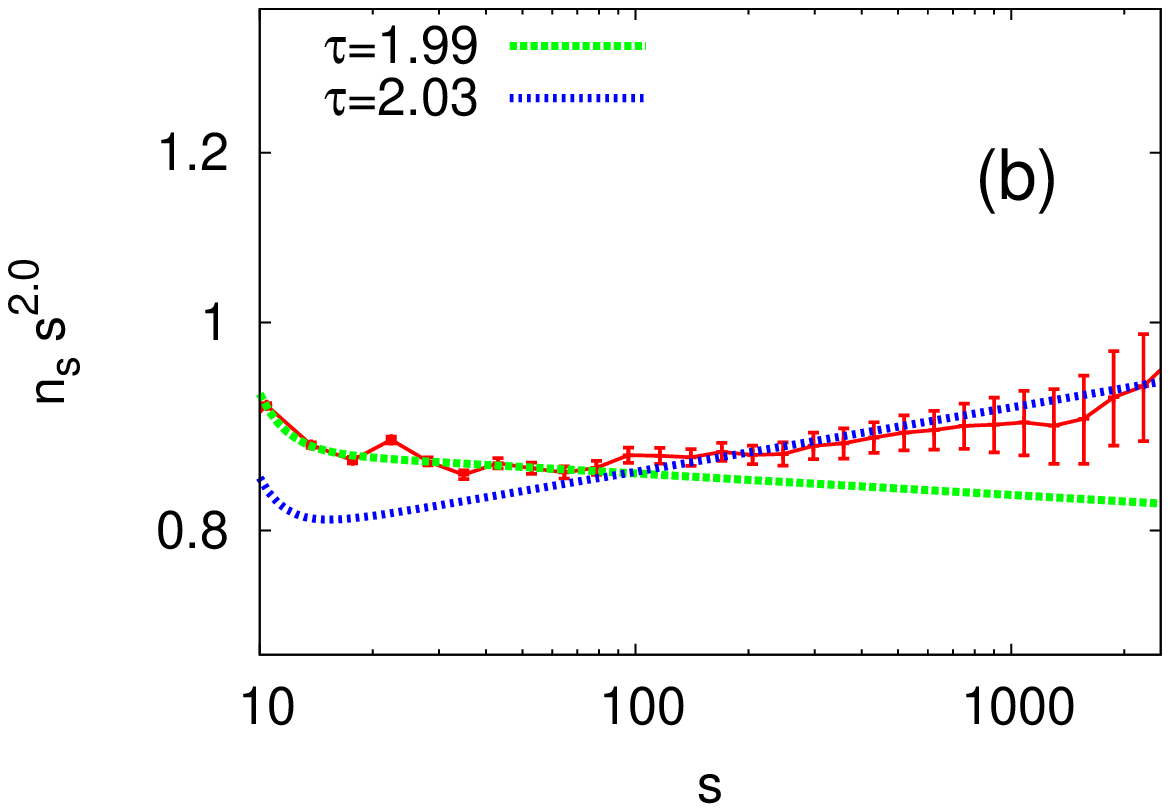}
\caption{(Color online) (a)
Cluster size distributions $n_s$ for the Ising model in
the $L \times L$ square lattices at the critical temperature $T_c$.
This figure is plotted in the form of $\ln (n_s s^\tau)$ versus
the cluster size $s$
with the predicted exponent $\tau = 1 + d/d_f = 379/187 \approx 2.03$.
The errorbars are of about the same sizes as the symbols for 
the scaling regime of $s \lesssim 1000$ and omitted for better visibility.
(b) Estimation of the Fisher exponent $\tau$ for $L=320$
(see text for details). Log-binned $n_s$ are shown in the form of
$n_s s^{2.0}$ versus $s$ (symbols) and the lines are results from the curve-fitting
to $n_s s^\tau \sim 1 + As^{-\Delta}$ with $\tau$ fixed to $2.03$ and $1.99$.
It is clearly seen that the correct value of $\tau$ is between 1.99 and 
and 2.03 and we thus conclude $\tau = 2.01(2)$. 
}
\label{fig:dist}
\end{figure}

\begin{figure}
\includegraphics[width=0.42\textwidth]{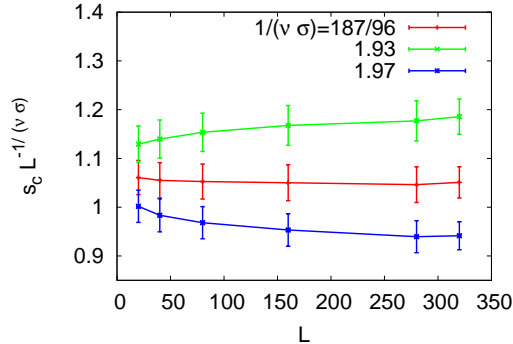}
\caption{(Color online)
Characteristic cluster size $s_c$ in the $L \times L$ square
lattices at $T_c \approx 2.27 (J/k_B)$.
The data points are scaled by $L^{-1/({\nu \sigma})}$ with varying
$1/{(\nu\sigma)}$ around $187/96 \approx 1.95$, so that all the data points
lie on a horizontal line within error bars. This estimate results in
$1/(\nu\sigma) = 1.95(2)$.
}
\label{fig:critical}
\end{figure}

\begin{figure}
\includegraphics[width=0.42\textwidth]{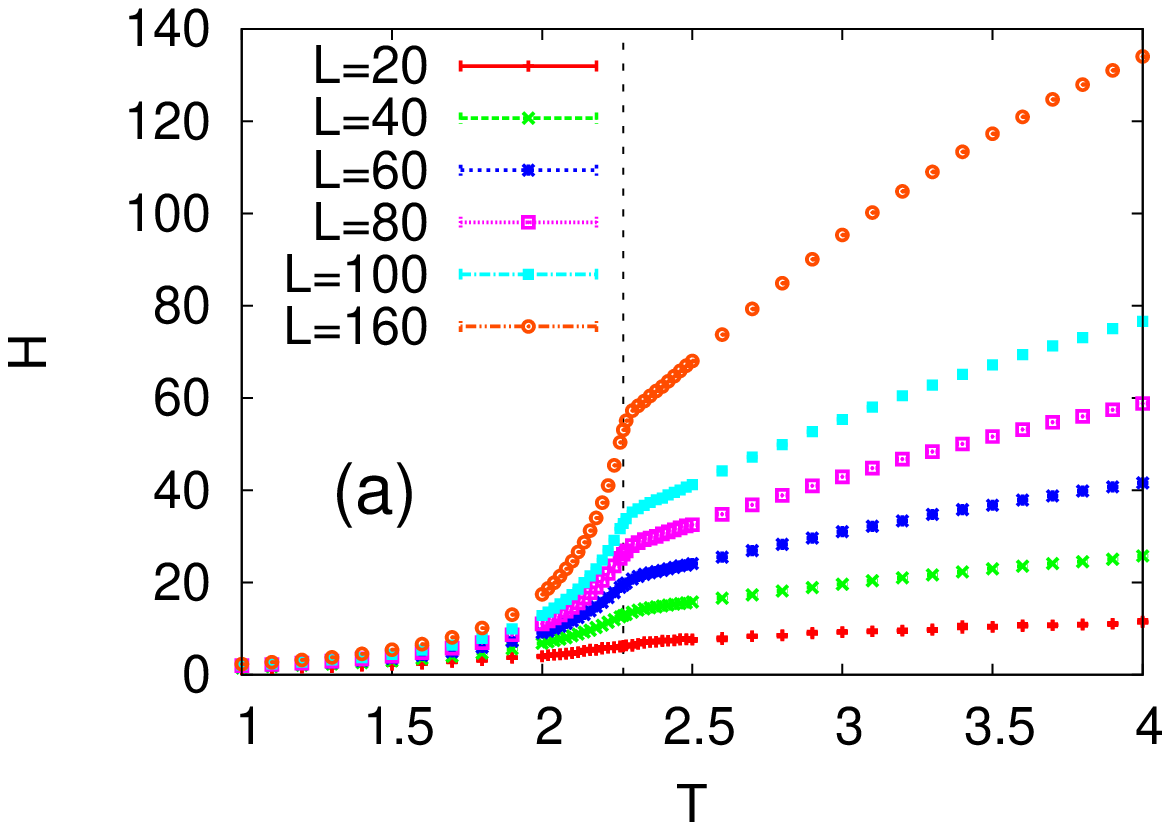}
\includegraphics[width=0.42\textwidth]{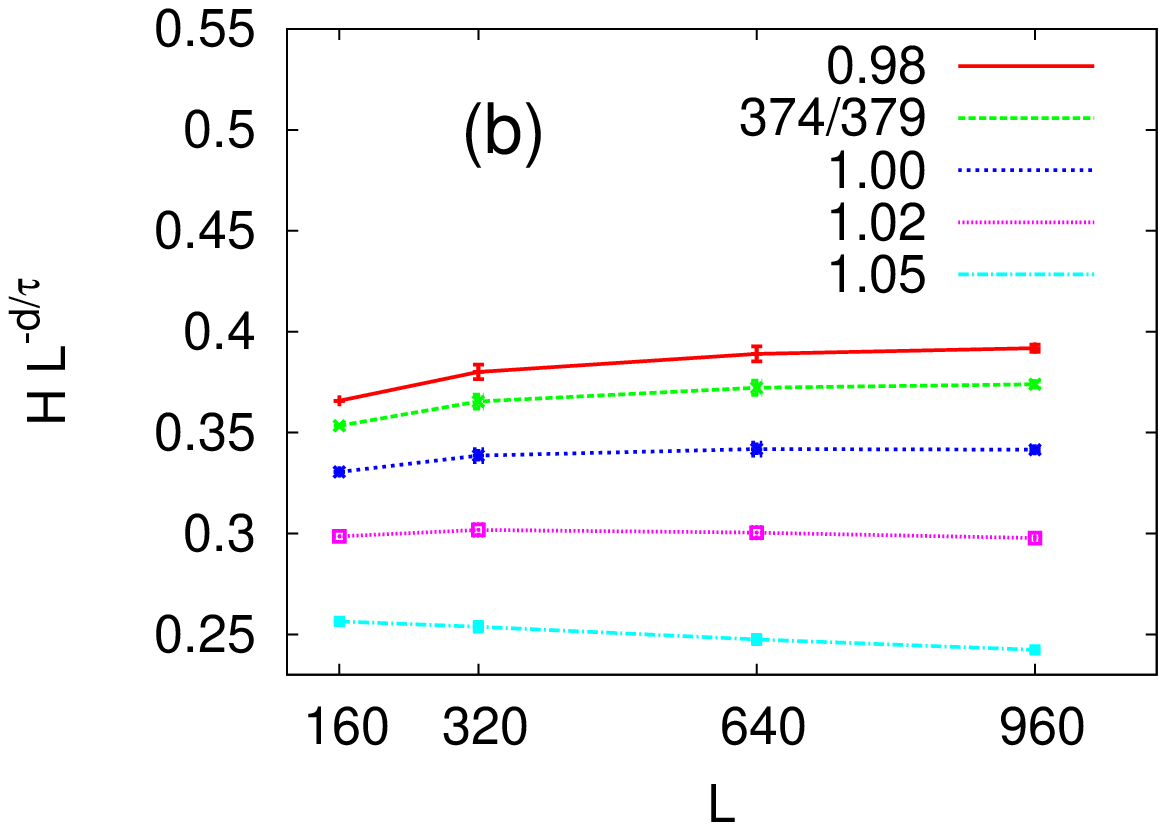}
\includegraphics[width=0.42\textwidth]{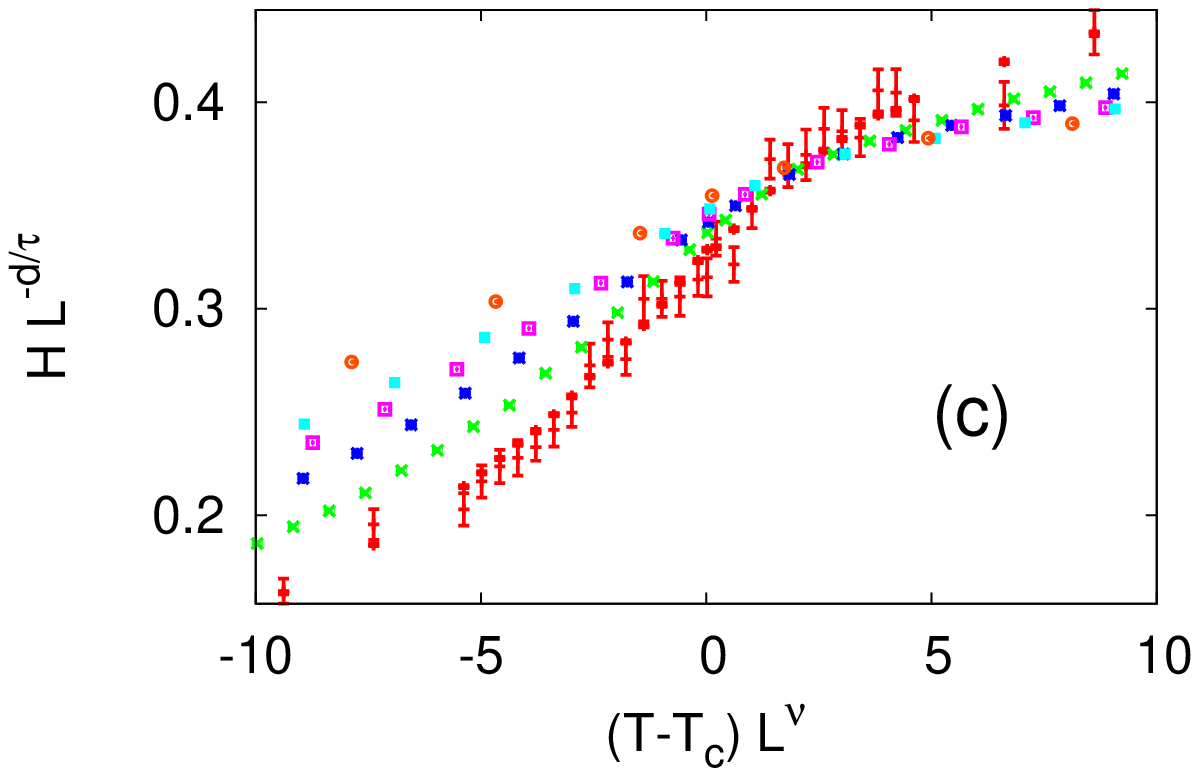}
\includegraphics[width=0.42\textwidth]{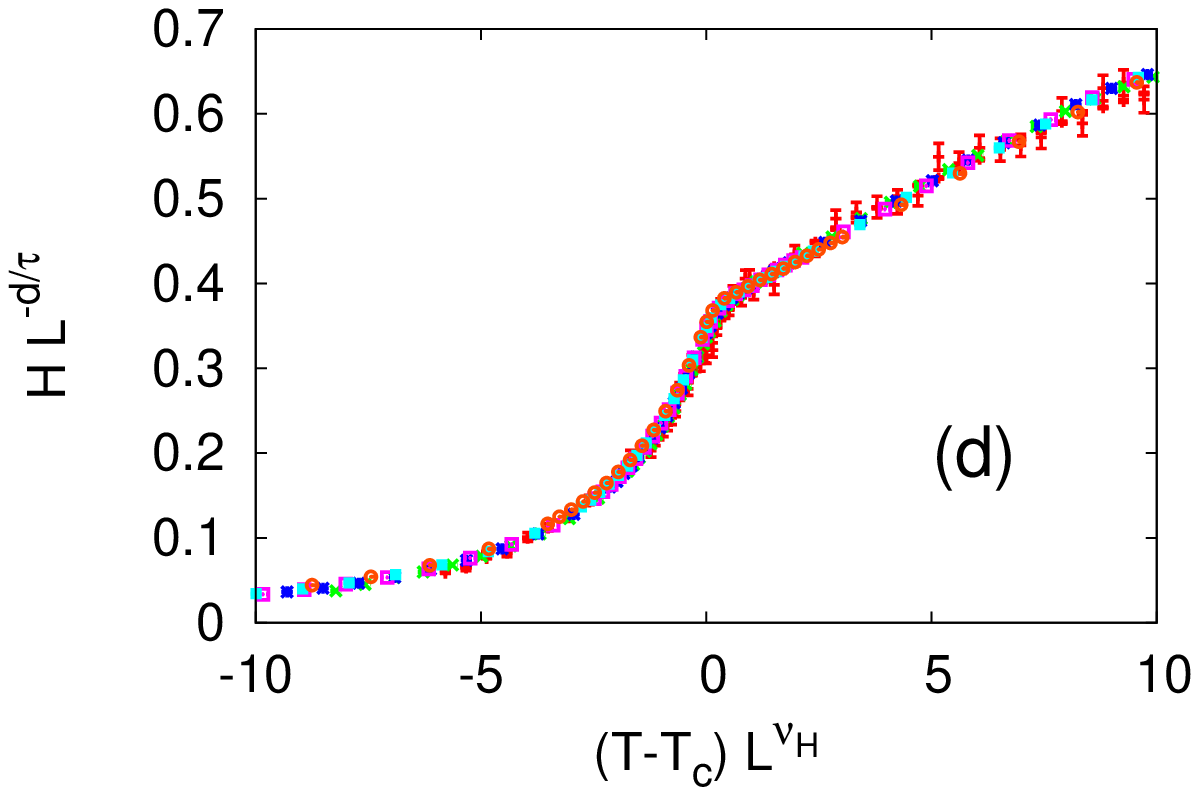}
\caption{(Color online) (a) Heterogeneity $H$ is drawn as a function of
$T$ for the 2D Ising model in the $L \times L$ square lattices. The dotted line
indicates $T_c$, and error bars are smaller than the data points.
(b) We obtain $X_H = 1.00(2)$ by scaling $H(t=0,L)$ as a function of
$L$ for the largest sizes at $T_c$.
(c) Scaling by using the conventional FSS ansatz in
Eq.~(\ref{eq:hetero_ansatzFSS}) with
$\nu = 1$ and $ X_H = 374/379\approx 0.99$.
(d) Scaling collapse using Eq.~(\ref{eq:hetero_FSS}) with
$\nu_H = 379/192 \approx 1.97$ and $ X_H = 374/379\approx 0.99$.
}
\label{fig:hetero}
\end{figure}

The cluster distribution in the 2D Ising model obeys the following
form~\cite{christensen} :
\begin{equation}
n_s =s^{-\tau} F_{n_s} \left(|t| s^x \right),
\label{eq:scaling_form_size}
\end{equation}
where $n_s$ is number of clusters with size $s$.
This scaling form is valid near $T_c$ and with
zero magnetic field. At $T=T_c$, it leads to
$n_s \sim s^{-\tau}$. The scaling relation Eq.~(\ref{eq:tau}) with 
$d_f=187/96$ for the 2D Ising model gives us $\tau=379/187 \approx 2.03$,
which is shown to be consistent with our numerical results as shown in
Fig.~\ref{fig:dist}(a). For a  more careful analysis, we allow the
correction to the scaling form as $n_s \sim s^{-\tau} (1 + As^{-\Delta} + \ldots)$ 
with a constant $A$ and a correction exponent $\Delta > 0$ and apply it 
for $L=160$ and $320$. We find that our numerical results are described sufficiently well by
$\tau = 2.01(2)$ as shown in Fig.~\ref{fig:dist}(b).
However, more precise estimation of $\tau$ is a formidable task, since it depends on the choice
of the scaling region of $s$ and also a further increase of $L$ can alter the estimation made
for smaller sizes. A better way is then to cross check with the outcomes of scaling relations 
as to be discussed below.

The peaks at the tail part in Fig.~\ref{fig:dist}(a)
are due to giant clusters,
exclusion of which yields the  monotonically decreasing distribution
without the peaks instead. This allows us to approximate
$F_{n_s}\left(|t| s^x \right)$ as an exponentially decaying function
$e^{-s/s_c}$ where
$s_c$ is the characteristic cluster size similar to the one used in
percolation~\cite{Noh}.
We may also identify $s_c$ with a peak position of $n_s$
near the tail in Fig.~\ref{fig:dist}(a),
and either way gives similar scaling
behavior, $s_c \sim |t|^{-1/\sigma}$.
In order to obtain the exponent $\sigma$, we apply the standard technique of
the FSS method to $s_c$:
\begin{equation}
s_c = L^{\frac{1}{\nu \sigma}} F_{s_c} \left(|t| L^{1/\nu} \right).
\label{eq:Sc_FSS}
\end{equation}
The exponent $\sigma$ can be then
obtained by analyzing simulation results with varying $L$ at $T = T_c$, by
which we
estimate the value of the exponent $1/{(\nu\sigma)} \approx 1.95(2)$
(see Fig.~\ref{fig:critical}).
Using the correlation-length
exponent $\nu = 1$ of the 2D Ising model, $1/\sigma
\approx 1.95(2)$ is numerically obtained.
The scaling relation $\sigma =
1/(\nu d_f)$ allows us to check our numerical results
with the predicted value $1/\sigma = 187/96 \approx 1.95$
as shown in Fig.~\ref{fig:critical}.
Again the agreement is compelling, and
our estimate favors this value over Suzuki's conjecture
$15/8 = 1.875$.

We next carry out the FSS analysis of $H$ in Fig.~\ref{fig:hetero}. 
First, we scale our numerical results [see Fig.~\ref{fig:hetero}(a)] using a
conventional FSS form:
\begin{equation}
H(t, L) = L^{X_H} F_H \left(t L^{1/\nu} \right),
\label{eq:hetero_ansatzFSS}
\end{equation}
with a scaling exponent $X_H$.
Although we can find the scaling exponent $X_H \approx 1.00(2)$ by
numerically observing $H(t=0,L)$ as a function of $L$
[see Fig.~\ref{fig:hetero}(b)],
we cannot observe scaling collapse with $\nu=1$ as shown in
Fig.~\ref{fig:hetero}(c).
Consequently, the argument of $F_H$ in the conventional FSS form
Eq.~(\ref{eq:hetero_ansatzFSS}) cannot be valid for $H$.
This implies also that
the correct scaling form should take into account the competition between
cluster sizes as in percolation.
By the same reasoning as above, a new scaling form is expected to be
\begin{equation}
H(t,L) = L^{d/\tau} F_H(t L^{1/\nu_H}),
\label{eq:hetero_FSS}
\end{equation}
where $\nu_H = \tau/(d\sigma)$.
Using $\tau = 379/187$ and $1/\sigma = d_f = 187/96$,
we expect the exponent to be $\nu_H =379/192 \approx 1.97$,
which indeed leads to a scaling collapse in a good quality
[see Fig.~\ref{fig:hetero}(d)].
In addition, our
numerical estimate $d/\tau = 1.00(2)$ contains the theoretical
prediction $374/379 \approx 0.99$ within the error bar.
We have obtained basically the same scaling behavior
for heterogeneity in the triangular lattice as well (not shown),
confirming its universality in the 2D Ising
model. Our results are summarized in Table~\ref{table:exp}.

\begin{figure}
\includegraphics[width=0.42\textwidth]{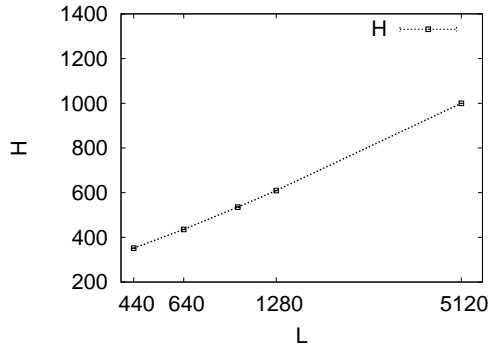}
\caption{Heterogeneity in the square lattices as a function of $L$, when
each site is assigned a random spin state between up and down at $T
\rightarrow \infty$.
Note that the horizontal axis is drawn on a logarithmic scale. Error
bars are shown but smaller than the data points.}
\label{fig:high}
\end{figure}

An interesting point in Fig.~\ref{fig:hetero}(a) is that $H$ keeps
increasing as $L$ grows even though $T$ is far higher than $T_c$.
Recall that all the correlation due to spin
interaction is destroyed in the infinite-$T$ limit, where we are
back to the percolation case. For the square lattice, the site-percolation
threshold is $p_c^{\rm square} = 0.59274602(4)$~\cite{ziff}, while the
infinite-$T$ Ising model corresponds to $p = 1/2$ due to the up-down symmetry.
Since Noh {\it et al.} have argued that $H
\sim \ln L$ when $|p-p_c| \gg L^{-1/\nu_H}$~\cite{Noh}, we expect that $H$
should have the logarithmic divergence for the infinite-$T$ Ising model,
which is confirmed by our numerical results (Fig.~\ref{fig:high}).
It is slower than the power-law divergence of $L^{d/\tau}$ at $T_c$,
so there will develop a peak at $T=T_c$ at a large $L$.
The existence of such a peak should be true for
the Ising model in the triangular lattice, too: As mentioned above, one can
find a percolating phase in the infinite-$T$ limit since $p_c^{\rm
triangle} = 1/2$.
The divergence at infinite $T$ will be therefore described by 2D percolation
exponents such as $H \sim L^{182/187}$, which is slightly slower than the
Ising case of $H \sim L^{374/379}$ at $T=T^{\rm triangle}_c = 4/\ln 3
\approx 3.64$~\cite{tri}. Returning back to the square
lattice, since we have a good reason to believe the existence of a peak at
$T=T_c$ for a large $L$, the monotonic shape of the scaling function in
Fig.~\ref{fig:hetero}(b) suggests that our observation might be still
subject to corrections to scaling. We have
indeed numerically found that the the logarithmic function $H(L)$ is hardly
distinguishable from linear increase when $L < \mathcal{O}(10^3)$, which
explains the increase of $H(L)$ in Fig.~\ref{fig:hetero}(a).

\begin{table}
\caption{Scaling exponents for heterogeneity of the 2D Ising model.
The numerical values are obtained by Monte Carlo calculations in this work,
whereas the analytic values are from tricritical exponents in the conformal
field theory.}
\begin{tabular*}{\hsize}{@{\extracolsep{\fill}}ccccc}\\ \hline \hline 
 & $d_f$ & $\tau$ & $d/\tau$ & $\nu_H$ \\ \hline
numerical & 1.95(2) & 2.01(2) & 1.00(2) & 1.96(5) \\
analytic & $187/96$ & $379/187$ & $374/379$ & $379/192$ \\
 & $\approx 1.95$ & $\approx 2.03$ & $\approx 0.99$ & $\approx 1.97$ \\
\hline \hline
\end{tabular*}
\label{table:exp}
\end{table}

In summary, we have shown that the cluster-size heterogeneity $H$ of the 2D
Ising model is scaled by the recently suggested FSS form using the exponent
$\nu_H$~\cite{Noh}, instead of the correlation-length exponent $\nu$.
The finite-size effects in measuring this quantity are quite substantial,
especially if compared to those in percolation. We have argued
the main reason that $H$ does not converge to a constant as the system size
increases when $T>T_c$ but still has weak divergence due to its geometric
nature. In spite of this difficulty, the scaling exponents at $T=T_c$ are in
nice agreement with theoretical predictions (see Table~\ref{table:exp}).
This justifies the validity of this observable as well as the use of $\nu_H$
instead of $\nu$ in the FSS analysis of the Ising system.

\acknowledgments
This work was supported by the National Research
Foundation of Korea (NRF) grant funded by the Korea government (MEST) (Grant
No. 2011-0015731).


\end{document}